\documentclass[showpacs,preprint,amssymb]{revtex4}
\usepackage{graphicx}
\usepackage{dcolumn}
\usepackage{bm}
\def\be{\begin{equation}} 
\def\ee{\end{equation}}
\def\bea{\begin{eqnarray}} 
\def\eea{\end{eqnarray}}
\def\line{\hbox to \hsize}    
\def\frac #1#2{{#1\over #2}}

\def\tr{{\rm  tr\,}}

\def \x{{\bf x}}

\def \dbrak #1#2{{\langle#1\vert#2\rangle}}

\def\1{\mbox{\bf 1}}
\def\bm#1{\mbox{\boldmath$#1$}} 

\begin{document}

\title{Berry Phase, Lorentz Covariance,  and Anomalous Velocity for Dirac and Weyl Particles}

\author{ MICHAEL STONE}

\affiliation{University of Illinois, Department of Physics\\ 1110 W. Green St.\\
Urbana, IL 61801 USA\\E-mail: m-stone5@illinois.edu}   

\author{VATSAL DWIVEDI}

\affiliation{University of Illinois, Department of Physics\\ 1110 W. Green St.\\
Urbana, IL 61801 USA\\E-mail: vdwived2@illinois.edu}   

\author{TIANCI ZHOU}

\affiliation{University of Illinois, Department of Physics\\ 1110 W. Green St.\\
Urbana, IL 61801 USA\\E-mail: tzhou13@illinois.edu}

\begin{abstract}  We consider the relation between   spin and the Berry-phase contribution to  the anomalous velocity of massive and massless  Dirac particles. We  extend the  Berry connection that  depends only on the spatial components of the particle momentum to one  that depends on the the space and time components in a covariant manner. We show that this covariant Berry connection captures  the Thomas-precession part of the Bargmann-Michel-Telegdi spin evolution, and contrast it with the traditional (unitary, but not naturally covariant) Berry connection that   describes  spin-orbit coupling.
We  then consider how the covariant connection  enters  the    classical relativistic dynamics of spinning particles due to  Mathisson, Papapetrou and Dixon.  We discuss the problems that arise with Lorentz covariance in the massless case,  and trace them mathematically to a failure of the Wigner-translation part of the massless-particle little group to be an exact gauge symmetry  in the presence of interactions, and physically to the fact that the measured position  of a massless spinning particle is  necessarily observer dependent.

\end{abstract}


\maketitle

\section{Introduction}

There has  been much recent interest   in the fluid dynamics of systems possessing anomalous conservation laws \cite{loga1,loga2,loga3,dubovsky1,dubovsky2}. An unexpected consequence of this work has been the discovery that anomalies, which are usually though of as being purely quantum mechanical  effects, can be     
extracted  from the classical kinetic theory of a degenerate  gas of Weyl fermions \cite{stephanov}. The  incompressibility of phase space   allows  the anomalous inflow    of  particles from the negative-energy Dirac sea into  the positive-energy Fermi sea \cite{haldane,son1,chen-pu-wang-wang}  to  be  reliably counted by keeping track of the density  flux   near the Fermi surface where a    classical  Boltzmann equation becomes sufficiently accurate  for this purpose. The only required  quantum input  is knowledge of how to normalize the phase space-measure  and the inclusion of a  Berry-phase  effect. The Berry phase causes the  velocity of the particle to  no longer be  parallel to its momentum.  Instead an   additional ``anomalous velocity'' appears as  a  momentum-space  analogue of the Lorentz force  in which the electromagnetic field tensor is replaced by the Berry curvature, and the particle velocity by $\dot {\bf k}$.  
The Berry phase also  alters the classical canonical structure so that ${\bf x}$ and ${\bf k}
$ are no longer conjugate variables,  and $d^3kd^3x$ is no longer the element of phase space volume 
\cite{niu,duval}. 

It is possible to extend these derivations to the non-abelian anomaly  \cite{stone-dwivedi}  and  to higher dimensions \cite{dwivedi-stone}, 
but  the kinetic-theory  used in all these papers is based on Hamiltonian dynamics where time and space are treated very differently. It is therefore a challenge to  to make the formalism  manifestly covariant so that a coupling to gravity might  be included.    Indeed it is not easy to see  how even flat-space Lorentz invariance is realized in the Hamiltonian kinetic theory.  This issue was raised in  \cite{son2} and the curious manner in which the dynamical variables must  transform  was made clear  in  \cite{son3}.

The most obvious problem with extending the $3$-dimensional Hamiltonian formulation to a covariant  3+1 version is that the  Berry curvature  is a differential form in only the three spatial components of the momentum. In a formalism  that treats space and time on an equivalent footing we would expect  the connection   to involve differentials of  all four components of the energy-momentum vector.  
 In this paper we show how to make such an  extension, and in doing so we make a connection between the Hamiltonian formalism with its Berry phase modification and the relativistic classical mechanics of spinning particles.  

In section \ref{SEC:covariant-berry} we use a WKB  solution to the massive-Dirac equation  to motivate an unconventional, but covariant,  Berry connection that  captures the geometric Thomas precession of the spin. We contrast the properties of this Berry connection with the traditional,   non-covariant  Berry connection  whose importance  in   the dynamics of charged  Dirac particles  was revealed   in \cite{bliokh-dirac,berard-mohrbach,niu-review}. 
 In section \ref{SEC:classical-spin} we introduce a   classical mechanical action for a spinning particle interacting with a gravitational field. This manifestly covariant action gives  rise to the well-known Mathisson-Papapetrou-Dixon equations \cite{mathisson,papapetrou,dixon}, and we show how these equations can be recast to make explicit  the role of the  covariant Berry connection.  In section \ref{SEC:massless} we   discuss the problems that arise when the particle mass becomes zero, and show how these arise from a hidden gauge invariance of the free action.  After selecting a natural gauge fixing condition, the covariant  action reduces to  the Berry-connection actions  used in  \cite{stephanov,stone-dwivedi,dwivedi-stone}.  Mathematically, it is the necessity of gauge fixing that is   responsible for the curious Lorentz transformation laws that appear in \cite{son3}, physically it is because the ``position'' of a massless spinning particle is an observer-dependent concept.      The  gauge invariance of the massless action  is only approximate in the presence of introduction of interactions and this leads to the  gauge-fixed action not being exactly equivalent to the manifestly covariant action. We argue   that this is  perhaps not surprising as in a massless system the adiabatic approximation that is tacit in any system involving a Berry connection can be violated by  a sufficiently large Lorentz transformation.  
 
 A discussion section  
 addresses the physical origin  of the anomalous velocity. Finally, several derivations  that would be intrusive in the main text  appear in  appendices \ref{SEC:appendix-spin}-\ref {SEC:appendix-centroid}.

\section{A Covariant Berry connection}
\label{SEC:covariant-berry}

That a   Berry phase  gives rise to an   anomalous velocity correction was first observed in the band theory of solids. We begin with a brief account of how the effect  appears  there, and why  a similar correction is expected in the  motion of Dirac particles.

 \subsection{Lorentz covariance versus  the  Berry phase}

 A semi-classical wave-packet analysis \cite{blount,sundaram-niu} shows that the motion of  a  charge-$e$ Bloch electron  in  an energy  band  in a crystalline solid  is  governed by the equations  
 \bea
\dot {\bf k} &=&  -\frac{\partial {\mathcal H}}{\partial {\bf x}}+  e(\dot {\bf x}\times {\bf B}), 
\label{EQ:lorentz}\\
\dot {\bf x}&=&\frac{\partial {\mathcal H} } {\partial  {\bf k}}  - \dot {\bf k} \times {\bm \Omega}.
\label{EQ:chang-niu}
 \eea
  The effective hamiltonian ${\mathcal H}=\varepsilon({\bf k})+e\phi({\bf x})$ includes the  band-energy $\varepsilon({\bf k}) $  as a function of the crystal momentum ${\bf k}$,  together with the  interaction with the  scalar  potential $\phi$. 
 The    vector  ${\bm \Omega}$ with components $\Omega_i= \textstyle{\frac 12}\epsilon_{ijk} \Omega_{jk}$ is a   Berry curvature that  accounts    for  the effects of  all other energy bands. The magnetic field ${\bf B}$ is a function of ${\bf x}$ only, and ${\bf \Omega}$ is a function  of ${\bf k}$ only. The  $-\dot {\bf k}\times {\bm \Omega}$ term in (\ref{EQ:chang-niu})  is  the  {\it anomalous velocity\/} correction to the na{\"i}ve group velocity $\partial \varepsilon/\partial{\bf k}$. This  correction  arises because  different momentum components of  a localized wave-packet accumulate different geometric phases when ${\bf k}$ is changing and the Berry curvature is non-zero   \cite{chong}. These ${\bf k}$-dependent  geometric phases are just as significant in determining the wave-packet position as  the ${\bf k}$-dependent dynamical phases  arising from  the dispersion equation $\omega=\varepsilon({\bf k})$. A nice illustration of the effect of the anomalous velocity on a particle trajectory is to be found in \cite{horvathy}.
 
 Now  a  Dirac Hamiltonian can be thought of a Bloch system with  two energy bands $\varepsilon({\bf k})\equiv  E({\bf k})=\pm \sqrt{{\bf k}^2+m^2}$, and  each  band possesses  a non-zero Berry curvature  \cite{bliokh-dirac,berard-mohrbach,niu-review}.  Consequently  (\ref{EQ:lorentz}) and  (\ref{EQ:chang-niu})  should also describe the semi-classical  motion  of a relativistic spin-$\textstyle{\frac 12}$ particle.  This raises an interesting issue.  We  expect that  the equation of  motion of a Dirac particle can be written   in a  manifestly  Lorentz invariant  form, but  it is not immediately obvious how to  massage the  Dirac version of (\ref{EQ:lorentz}) and (\ref{EQ:chang-niu}) into  covariant expressions.   The first line (\ref{EQ:lorentz})  can be written as $\dot k_\mu = eF_{\mu\nu}\dot x^\nu$, but for (\ref{EQ:chang-niu}) how does   one define a 3+1 dimensional  analogue of the Maxwell tensor $F^{\mu\nu}$ for the intrinsically three-dimensional Berry curvature $\Omega_{ij}$?  
 
 \subsection{Covariant WKB approximation for  the Dirac equation}
 
In order to obtain a manifestly Lorentz invariant semi-classical equation of motion for a Dirac particle, we need to extend the   non-covariant Berry connection  to one in which space and time components are treated equally. Now the simplest semi-classical approximation to any wave equation is that of WKB.   We therefore construct a  WKB approximation to the Dirac equation coupled to an  externally imposed Maxwell field.  We maintain covariance at each step,  anticipating that a covariant version of  Berry curvature will play some role.  WKB approximations to the Dirac equation have a long history, going back to   W.~Pauli in 1932 \cite{pauli}. More recent references are \cite{rosen,rubinov,bolte}.  None of these works make use of the particular covariant approach  that we introduce   here.

We take the particle to have   charge $e$ (a positive number when  the charge is positive)  and to have positive mass 
$m$.
Let  $x^\mu=(t,{\bf x})$, and  seek a positive-energy  WKB solution 
\be
\psi(x)= a(x) e^{-i\varphi(x)/\hbar}, \quad a=a_0+\hbar a_1+\hbar^2 a_2+\ldots 
\ee
to the
Dirac equation 
\be
\left(i\hbar \gamma^\mu (\partial_\mu +ieA_\mu/\hbar)-m\right)\psi=0.
\label{EQ:dirac-for-wkb}
\ee
Here $\{\gamma^\mu,\gamma^\nu\}=2\eta^{\mu\nu}$ with  Minkowski metric $\eta^{\mu\nu}= {\rm diag}(+,-,-,-)$, and   $A_\mu=(\phi, -{\bf A})$.
 
Setting      $p_\mu\stackrel{\rm def}{=}  \partial_\mu \varphi=(E,-{\bf p})$, we have  at order $\hbar^0$
\be
\left(\gamma^\mu (p_\mu-eA_\mu)-m\right)a_0=0.
\label{EQ:dirac-hbar0}
\ee
We satisfy (\ref{EQ:dirac-hbar0})  by setting
$a_0(x)= u_\alpha(k(x)) C^\alpha(x)$ with $k_\mu= p_\mu-eA_\mu$ being the gauge-invariant kinetic momentum, and $u_\alpha(k)$ being a complete set of eigenspinor solutions to 
\be
(\gamma^\mu k_{\mu}-m)u_\alpha =0.
 \label{EQ:dirac-constant}
 \ee
In this equation the  kinetic four-momentum  $k_\mu=(E,-{\bf k})$ lies  on  the positive-energy mass shell: $E^2={\bf k}^2+m^2$, $E>0$.  We take the eigenspinors to have the covariant normalization $\bar u_\alpha u_\beta= \delta_{\alpha \beta}$. (See appendix \ref{SEC:appendix-spin} for details)
 
At order $\hbar^1$ we have
\be
\left(\gamma^\mu k_\mu - m\right)a_1+ (i\gamma^\mu \partial_\mu)a_0=0.
\ee
Now  if 
\be
\left(\gamma^\mu k_\mu -m\right)u_\alpha=0,
\ee
then 
\be
\bar u_\alpha \left(\gamma^\mu k_\mu -m\right)=0.
\ee
We can therefore eliminate  the influence of the unknown coefficient $a_1$ and deduce that
\be
\bar u_\beta\gamma^\mu\partial_\mu a_0=\bar u_\beta\gamma^\mu\partial_\mu (u_\alpha C^\alpha)=0.
\label{EQ:transport}
\ee
Equation (\ref{EQ:transport})  tells us how both the amplitude and spin components evolve   along the classical trajectory. We rewrite (\ref{EQ:transport}) as 
\be
\bar u_\beta\gamma^\mu u_\alpha ( \partial_\mu C^\alpha) + (\bar u_\beta\gamma^\mu\partial_\mu u_\alpha) C^\alpha =0,
\ee
and then use   (\ref{EQ:V-from-u})   from appendix \ref{SEC:appendix-general} to write  
\be
\bar u_\beta \gamma^\mu u_\alpha = \delta_{\alpha\beta} \frac{k^\mu}{m} \equiv \delta_{\alpha\beta} V^\mu,
\ee
and so  express the transport equation (\ref{EQ:transport}) as
\be
\left[\delta_{\alpha\beta}V^\mu \frac{\partial}{\partial x^\mu} +M_{\alpha\beta}\right] C^\beta =0.
\ee
Here
\be
M_{\alpha\beta}=\bar u_\alpha\gamma^\mu \frac{\partial}{\partial x^\mu}u_\beta,
\ee
and  $V^\mu= \gamma(1,  {\bf v})=k^\mu/m$ is the four-velocity corresponding to the ray-tracing group velocity 
\be
 {\bf v}= \frac{\partial E}{\partial {\bf k}}.
\ee
Thus  the combination 
\be 
V^\mu \frac{\partial}{\partial x^\mu} \equiv \frac{d}{d\tau}
\ee
is a   convective derivative with respect  to  proper time along the particle's trajectory.  The $({\bf x},{\bf k})$ trajectory itself is given by Hamilton's ray-tracing  equations and coincides with that of a spinless charged particle in the background field.  There  is no sign of the anomalous velocity. As pointed out in \cite{rubinov}, this   absence  is to be expected  because  both the intrinsic spin and magnetic moment of a Dirac particle are  proportional to $\hbar$, and vanish in the classical limit. Thus  leading-order  WKB  is not  able to account for the effect of the spin on the particle's motion.  Nonetheless  the  {\it ratio\/}  of the magnetic moment to the spin  angular momentum  is independent of $\hbar$.   As a consequence     leading-order  WKB  {\it is\/} adequate for obtaining the Bargmann-Michel-Telegdi (BMT) equation \cite{BMT} that describes the   effect of the magnetic field on the spin evolution. A  Berry connection  is a key  ingredient in  this equation.

To isolate the Berry connection, we decompose 
\be
M_{\alpha\beta}= \frac{1}{2} (M_{\alpha\beta}+M_{\beta\alpha}^*)+\frac{1}{2} (M_{\alpha\beta}-M_{\beta\alpha}^*),
\ee
and,  from   equation (\ref{EQ:V-from-u}), recognize that   
\be
\frac{1}{2} (M_{\alpha\beta}+M_{\beta\alpha}^*)= \frac 12  \frac{\partial V^\mu}{\partial x^\mu}.
\ee
We now insert  the completeness relation ${\mathbb I}=u_\lambda\bar u_\lambda-v_\lambda \bar v_\lambda$ as intermediate states in the definition of $M_{\alpha\beta}$. From the positive-energy $u_\lambda\bar u_\lambda$ terms we get
\be
(\bar u_\alpha \gamma^\mu u_\lambda)\left( \bar u_\lambda  \bar \frac{\partial}{\partial x^\mu}u_\beta\right) = \left(V^\mu \frac{\partial k^\nu }{\partial x^\mu} \right)\left(\bar u_\alpha \frac{\partial}{\partial k^\nu} u_\beta\right)= -i {\mathfrak a}_{\alpha\beta,\nu} \frac{d k^\nu}{d\tau}.
\ee
The quantity 
\be
{\mathfrak a}_{\alpha\beta,\nu}\stackrel{\rm def}{=} i\bar u_\alpha \frac{\partial u_\beta}{\partial k^\nu}
\ee
is an unconventional  Berry-phase-like  connection. It is  unitary  only with respect the non-positive-definite inner product $\dbrak{\psi}{\chi}\equiv \psi^\dagger \gamma^0 \chi$, but makes use of all four components of $dk^\nu$ and  is constructed out of Lorentz-covariant objects. We will therefore refer to it as the {\it covariant Berry connection\/}. 

 The  contribution of the negative energy intermediate states  $-v_\lambda \bar v_\lambda$ is an example of \\ Littlejohn's  ``no-name'' phase \cite{littlejohn}. After some labour,  we find that their contribution is 
 \be
-\frac 12 \left(\bar u_\alpha\gamma^\mu v_\lambda \bar v_\lambda \frac{\partial}{\partial x^\mu}u_\beta -(\alpha\leftrightarrow \beta)^*\right) = \frac{ie}{2m} S_{\alpha\beta}^{\mu\nu} F_{\mu\nu}C^\beta,
 \ee
 where $\dot k^\nu= dk^\nu/d\tau$, 
\be
(S_{\mu\nu})_{\alpha\beta}= \bar u_\alpha \left( \frac{i}{4}[\gamma_\mu,\gamma_\nu]\right)u_\beta,
\label{EQ:S-definition}
\ee
and we have used  $k_\mu= \partial_\mu\varphi-eA_\mu$ to write
\be
\partial_\mu k_\nu-\partial_\nu k_\mu= -eF_{\mu\nu}.
\ee
The combined contribution of both sets of  intermediate states therefore leads to 
\be
\left[ \delta_{\alpha\beta}\left(V^\mu \frac{\partial}{\partial x^\mu}+ \frac 12 \frac{\partial V^\mu }{\partial x^\mu} \right)-i\dot k^\nu ({\mathfrak a}_\nu)_{\alpha\beta} +
\frac{ie}{2m} S_{\alpha\beta}^{\mu\nu} F_{\mu\nu}\right]C^\beta=0.
\label{EQ:covariant-transport}
\ee
The divergence of the four-velocity in (\ref{EQ:covariant-transport}) accounts for the change in amplitude due to geometric focussing. The remaining terms describe how the spin evolves  through  its interaction with the external field, and as a result of its parallel transport under  the Berry connection.  

The combination  $ S^{\mu\nu} F_{\mu\nu}$ is Lorentz invariant, so we can evaluate it in the particle's rest frame where 
\be
\left(\frac{e}{2m}\right) (S^{\mu\nu})_{\alpha\beta} F_{\mu\nu}\to -\left( \frac{e}{m}\right) {\bf B}\cdot \left(\frac {{\bm \sigma}}{2}\right)_{\alpha\beta}.
\label{EQ:larmor}
\ee
Since the unitary operator for a   rotation at angular velocity ${\bm \omega}$ is 
 $U(t)=\exp\{-i {\bm \omega}\cdot ({\bm \sigma}/2)t\}$ we see that (\ref{EQ:larmor})  accounts for the Larmor precession ${ \bm \omega}_{\rm Larmor} =- |{\bm  \mu}| {\bf B}$  of the spin due to its ${\bm \mu}=(e/m){\bf S}$ Dirac-value magnetic moment.  The two-by-two matrix  $(e/2m) {\bf B} \cdot {\bm \sigma}$ acts on  the {\it polarization spinor\/}   $\chi_\alpha$ that is defined in    (\ref{EQ:chidef}).  Polarization  is  the spin  measured   in the  rest frame of the particle \cite{rosen}. 
 
To understand the origin   of the Berry connection term we  use the explicit formul{\ae}\  for $u_\alpha({\bf k})$ given in (\ref{EQ:u-alpha-covariant})   to evaluate  
\bea 
{\mathfrak a}_{\alpha\beta, \nu} \dot k^\nu &=&\frac{1}{m^2(1+\gamma)} ({\bf k}\times \dot {\bf k})\cdot \left(\frac {\bm \sigma}{2}\right)_{\alpha\beta} \nonumber\\
&=& \frac{\gamma^2}{1+\gamma}({\bm \beta}\times \dot {\bm \beta})\cdot \left(\frac {\bm \sigma}{2}\right)_{\alpha\beta} \nonumber\\
&=& - {\bm \omega}_{\rm Thomas} \cdot \left(\frac {\bm \sigma}{2}\right)_{\alpha\beta}.  
\label{EQ:cov-connection}
\eea
 Here ${\bm \beta}\equiv {\bf k}/E$, and 
 \be
 {\bm \omega}_{\rm Thomas}= - \left(\frac{\gamma^2}{1+\gamma}\right) ({\bm \beta}\times \dot {\bm \beta})\
 \ee
 is a standard  expression  for the  Thomas-precession angular velocity.  Our  covariant  Berry transport   is therefore nothing other than Thomas precession --- 
{\it i.e.\/}\   parallel transport on the tangent  bundle of  the positive-mass hyperboloid embedded in Minkowski-signature momentum-space \cite{rhodes}. The minus sign occurs because the mass-shell hyperboloid is  a  negative-curvature Lobachevskii space. 

The matrix-valued connection one-form is defined by 
\be
{\mathfrak a} \stackrel{\rm def}{ =}\frac{1}{m^2(1+\gamma)} \left(\frac {\bm \sigma}{2}\right)\cdot ({\bf k}\times  d{\bf k}),
\ee
and the associated matrix-valued curvature  ${\mathfrak F}= d{\mathfrak a}-i {\mathfrak a}^2$ is  
\be
{\mathfrak F}=  \frac{1}{2m^2 \gamma}\left\{ \frac 12  \left( {\bm \sigma}+ \frac{ ({\bf k}\cdot {\bm \sigma}){\bf k}}{m^2(1+\gamma)}\right)\right\}\cdot (d{\bf k}\times d{\bf k}).
\label{EQ:3d-cov-curvature}
\ee
 The connection-form and the curvature do not \underline{look} covariant as they  involve only the spatial components of $dk^\mu$.  This is  a consequence of the way we wrote  $u_\alpha({\bf k})$ in (\ref{EQ:u-alpha-covariant}). In appendix \ref{SEC:appendix-general} we avoid explicit formal{\ae}\  for $u_\alpha$ and  use only general properties of the Dirac equation to  obtain  an expression for  the  curvature in arbitrary dimensions. We  find that 
 \be
{\mathfrak F}_{\alpha\beta}\equiv (d{\mathfrak a}-i {\mathfrak a}^2)_{\alpha\beta} = \frac{1}{2m^2} (S_{\mu\nu})_{\alpha\beta}\,dk^\mu\wedge dk^\nu,
\ee
where $(S_{\mu\nu})_{\alpha\beta}$ was defined in equation (\ref{EQ:S-definition}).
This form of the curvature is manifestly covariant and  contains both space and time components of $dk^\mu$. The $dk^\mu$ are not independent however, but are constrained by the mass-shell condition $k^2=m^2$. If we desire, therefore, we may   eliminate 
$dk^0$ as $dk^0= d\sqrt{{\bf k}^2+m^2}= k^i dk^i/E= -k_i dk^i /E$ and find 
\be
{\mathfrak F}_{\alpha\beta}=  \frac{1}{2m^2} \left( S_{ij}- \frac{k_i}{E} S_{0j} - S_{i0}\frac{k_j}{E}\right)_{\alpha\beta} dk^i \wedge dk^j,
\ee
where $i,j$ run over  space indices only. Evaluation of the required $S_{\mu\nu}$  matrix elements confirms that this reduced expression coincides with  (\ref{EQ:3d-cov-curvature}).
The combination of spin components in parentheses on  the right-hand side is a general-dimension  analogue of the space part of the  (3+1)-dimensional Pauli-Lubansky vector. We will therefore refer to it as the Pauli-Lubanksy tensor  (It is tensor only under space rotations. It is  not a Lorentz tensor). It will appear frequently in the rest of the paper and its geometric and physical significance is  further discussed in  appendices  \ref{SEC:appendix-spin} and \ref{SEC:appendix-centroid}. 

To verify  that parallel transport {\it via\/} the covariant Berry connection is nothing other than Thomas precision,   we show in appendix \ref{SEC:appendix-fermi-walker} that under such transport ({\it i.e.\/}\ no external torque or Larmor precession) the WKB approximation to the Dirac-field angular momentum  tensor $S_{\mu\nu} = \bar \psi (i [\gamma_\mu,\gamma_\nu]/4) \psi $ obeys 
\be 
\frac{ \partial S^{\mu\nu}}{\partial \tau} +    V^\nu \frac{\partial  V^\lambda}{\partial \tau} {S^\mu}_\lambda + V^\mu \frac{\partial V^\lambda}{\partial \tau} {S_\lambda}^\nu=0.
\label{EQ:fermi-walker}
\ee
Since   (\ref{EQ:nice-condition}) tells us that  $V^\mu S_{\mu\nu}=0$, and so (\ref {EQ:fermi-walker}) states that $S_{\mu\nu}$ is Fermi-Walker transported along the particle trajectory.   Thomas precession is simply  the evolution under Fermi-Walker transport of  vectors (such as the spin four-vector ${\bf S}$) that are   perpendicular to the four-velocity vector.  

\subsection{Comparison with the non-covariant  WKB approximation}

The traditional  form of the WKB transport equation is  obtained by expanding  $\psi= u_{\alpha}({\bf k})  K^\alpha(x)$ where the $u_{\alpha}$ are given the \underline{non-covariant} normalization $u^\dagger_\alpha  u_\beta=\delta_{\alpha\beta}$, and paired with negative-energy  solutions that, in terms of the covariant  $v_\alpha$, are given  $\gamma^{-1} v_\alpha(-{\bf k})$. These  non-covariant  spinors  have completeness relation ${\mathbb I}=u_\alpha u^\dagger_\alpha + v_\alpha v^\dagger_\alpha$. On using them as intermediate states we obtain  the alternative form of   transport equation found in  \cite{rosen,bolte}:
\be
\left\{\delta_{\alpha\beta} \left(\frac{d}{dt}+\frac 12 {\rm div}\, {\bf v}\right) +N_{\alpha\beta}\right\}C^\beta=0.
\ee
Here $t$ is the lab-frame time, ${\bf v}={\bm \beta}$ is the three velocity, and 
\bea
N_{\alpha\beta}&=& -i { a}_{\alpha\beta,i}\dot k^i  -i\left(\frac{e}{m}\right) {\bf B}\cdot \left({\bm \sigma}+ \frac 1{m^2} \frac{({\bf k}\cdot {\bm \sigma}){\bf k}}{\gamma+1}\right)_{\alpha\beta}\frac 1{2\gamma^2}\nonumber\\
&=& -i  { a}_{\alpha\beta,i} k^i  -i\left(\frac{e}{m}\right) \frac 1 {\gamma^2} {\bf B}\cdot( {\bf S}_{\rm lab})_{\alpha\beta}.
\eea
The  term  with the magnetic field ${\bf B}$ is again a  ``no-name'' phase that  arises from the the  negative-energy intermediate states   \cite{littlejohn}.
The Berry connection $a_{\alpha\beta,i}$ is here of conventional form
\bea
a_{\alpha\beta,i}dk^i  &\stackrel{\rm def}{=}& i u^\dagger_\alpha \frac{\partial u_\beta}{\partial k^i} dk^i   \nonumber\\
      &=& - \frac{\gamma}{1+\gamma}({\bm \beta}\times d{\bm \beta})\cdot \left(\frac {\bm \sigma}{2}\right)_{\alpha\beta}. 
\label{EQ:niu-connection}
\eea
Compared to the covariant Berry connection, (\ref{EQ:niu-connection})  lacks one power of  $\gamma$. More importantly, it has the {\it opposite sign}. The associated matrix-valued curvature is \cite{niu-review}
\bea
{\mathcal F}&=&  da-ia^2\nonumber\\
&=& - \frac 1{4m^2\gamma^3}\left({\bm \sigma}+ \frac 1{m^2} \frac{({\bf k}\cdot {\bm \sigma}){\bf k}}{\gamma+1}\right)\cdot (d{\bf k}\times d{\bf k}).
\label{EQ:niu-curvature}
\eea 
Again compared to the covariant expression ${\mathfrak F}$, the new curvature ${\mathcal F}$  lacks two powers of $\gamma$, and again has  the opposite sign. 
  
Both the covariant and the non-covariant transport equation lead to the same  BMT equation, but there is a different  distribution between terms  of  the dynamical Larmor precession and the geometric parallel transport.  In the covariant formulation we  have precession of the rest-frame  polarization ${\bf s}$ due to the magnetic field as seen by the particle in its \underline{rest frame}, and augmented   by the  geometric Thomas precession factor. This is how the BMT equation  is broken up in Jackson \cite{jackson},  
in his equation (11.166).   In the non-covariant formulation we have precession of the same rest-frame  polarization  ${\bf s}$,  but now due to the magnetic field as seen by  the spin in the \underline{lab frame} and   augmented  by the conventional Berry transport term. This is how the BMT equation is decomposed in  \cite{niu-review}, where the  connection (\ref{EQ:niu-connection}) and curvature (\ref{EQ:niu-curvature}) are obtained from a wave-packet approach.

The difference in sign between the two connections is accounted for by the different physical effects  that they capture.  The covariant connection  provides   the purely geometric Thomas precession effect.  The  non-covariant Berry connection  implements the spin-orbit coupling due to  the particle's motion viewed from  the lab frame \cite{mathur}. As  was  famously explained by Llewellyn Thomas \cite{thomas},  this spin-orbit coupling comes from two competing effects:  firstly  the Lorentz transform of the external field
that leads to the motion through an ${\bf E}$ field being  perceived as a ${\bf B}$ field, and secondly the Thomas precession  that half-undoes the Lorentz transformation contribution. The net  precession rate   therefore has opposite sign to   its Thomas-precession component.


\section{Classical  motion of particles with spin}  
\label{SEC:classical-spin}

 Rather than attempt to extend the WKB approximation to higher order, we will use symmetry consideration to construct  a  Hamiltonian action-principle  version of the dynamics that is manifestly covariant,  gives the  same  spin transport as the WKB approximation, but also  gives us an anomalous-velocity correction. 
 As our ultimate goal is   to understand  the effect of  gravity on the particle, we will from the outset take  our space-time to be curved.

 \subsection{Mathisson-Papapetrou-Dixon equations}

  There is an extensive    literature on the relativistic  classical dynamics of particles with spin, but a desire to  make contact with the Berry phase methods of \cite{stephanov,stone-dwivedi,dwivedi-stone}  suggests that we follow the  particular approach of \cite{kunzle,balachandran,stern}  and   take as our dynamical degrees of freedom  the position $x\in M$ (where $M$ is the $d$-dimensional space-time manifold)  and a vielbein frame $\tilde {\bf e}_a$ with $\tilde e_a^\mu \tilde e_b^\nu g_{\mu\nu}= \eta_{ab}$ where  $\eta_{ab}= {\rm diag}(+, -,-,\ldots, -)$. Our  phase space 
is then  the total space $P$ of a  Lorentz-frame bundle $\pi:P\to M$ equipped  with local coordinates  $(x^\mu, \tilde e_a^\mu)$ and  structure group  ${\rm SO}(1,d-1)$.  
 
It is convenient to introduce   a {\it reference vielbein\/} ${\bf e}_a$, again  with $e_a^\mu e_b^\nu g_{\mu\nu}= \eta_{ab}$.  This reference frame allows us to write    
\be
\tilde {\bf e}_a= {\bf e}_b {\Lambda^b}_a, \quad \Lambda\in {\rm SO}(1,d-1), 
\ee
and so equivalently  regard the dynamical degrees of freedom to be $x\in M$ and the Lorentz transformation $\Lambda$. 

We  assume  that  the space-time $M$ is equipped with a Riemann connection---and hence with covariant derivatives $\nabla_\mu$.    The reference vielbein then   defines  the   components of the spin connection 
${\omega^a}_{b\mu}$ by 
\be
\nabla_\mu {\bf e}_a= {\bf e}_b\,  {\omega^b}_{a\mu}.
\ee
We use these components to assemble  the spin-connection  one-form
\be
{\omega^a}_b = {\omega^a}_{b\mu}dx^\mu,
\ee
which lives  on the base-space M.  The associated Riemann curvature is  the base-space  two-form
 \be
 {R^a}_b= d{\omega^a}_b + {\omega^a}_c \wedge {\omega^c}_b.
 \ee 
As with any frame bundle, the connection on the base space automatically provides a decomposition of the tangent space at each point $p$ in the total space $P$ of the bundle into horizontal and vertical subspaces:  $T(P)= H\oplus V$.

We begin with particles with a non-zero mass $m$ and orient the frame so that  ${k}\equiv m \tilde {\bf e}_0$ is  the 4-momentum.  Thus $k^b= m {\Lambda^b}_0$ are the vielbein components of the momentum and $ k^\mu= m e^\mu_a {\Lambda^a}_0$ are its  coordinate components.
We also introduce a co-frame of one-forms 
\be
 {\bf e}^{*a}= e_\mu^{*a}dx^\mu 
\ee
where ${\bf e}^{*a}({\bf e}_b)\equiv  { e}^{*a}_\mu e^\mu_b= \delta^a_b$  and 
$
 e_\mu^{*a} =g_{\mu\nu}\eta^{ab}  e^\nu_b.
$
We then  set $\tilde{\bf e}^{*a}= {(\Lambda^{-1})^a}_b {\bf e}^{*b}$.
With  our $\eta_{ab}={\rm diag}(+,-,-,\ldots,-)$ signature we have $m\tilde {\bf e}^{*0}= k_\mu dx^\mu=k_a {\bf e}^{*a}$.

In \cite{stone-dwivedi,dwivedi-stone}, the action integral was written in terms of traces over some  faithful representation of the  spin or gauge  groups.  In the present case  we  could  use any faithful representation of the Lorentz group,   but it seems  natural  to make use of  Dirac matrices $\gamma_a$ and the Dirac representation $\Lambda\mapsto  D(\Lambda)$ that acts on them as 
\be
D(\Lambda)\gamma_aD(\Lambda^{-1})=  \gamma_b {\Lambda^b}_a.
\ee
We will simplify the notation by setting  $\lambda=D(\Lambda)$. 
In this section we use  the matrices 
\be
\sigma_{ab}= \frac 1{4} [\gamma_a,\gamma_b] 
\ee
as the Lorentz generators.
These matrices  obey 
\be
[\sigma_{ij},\sigma_{mn}]=\eta_{jm}\sigma_{in}  -\eta_{im}\sigma_{jn}-\eta_{jn}\sigma_{im}+\eta_{in}\sigma_{jm},
\ee
and 
\be
[\sigma_{ab},\gamma_c]= \gamma_a \eta_{bc}- \gamma_b \eta_{ac}.
\ee
We also have 
\be
{\rm tr}\{ \sigma_{ab}\sigma_{cd}\} =-\frac 14 {\rm tr}({\mathbb I})(\eta_{ac}\eta_{bd}- \eta_{ad}\eta_{bc}).
\ee

The covariant derivative acting on a spin field is
\be
\nabla_\mu \psi = \left(\frac{\partial}{\partial x^\mu}+\frac 1 2\sigma_{ab}\,{\omega^{ab}}_\mu\right)\psi,
\ee
and, as usual, we  regard the spin connection in the Dirac representation 
\be
\omega\equiv \frac 12\sigma_{ab}\,{\omega^{ab}}_\mu dx^\mu
\ee
as a matrix-valued one-form.

We can use the Lorentz transformation matrix $\lambda$  to   write
\be
k_a= \tr\{\kappa \lambda^{-1} \gamma_a \lambda\},
\ee
where $\kappa = m\gamma^0/{\rm tr}({\mathbb I})$. 
Similarly, we define a classical spin angular-momentum  tensor
\be
S_{ab}= {\rm tr}\{\Sigma\lambda^{-1}\sigma_{ab}\lambda\},
\ee
where  $\Sigma= \textstyle{\frac 12} \Sigma^{ab} \sigma_{ab}$.  

The quantities   $k_a$ and $S_{ab}$  are the true dynamical variables of the system. They are coordinates on the orbit of   $\kappa$ and $\Sigma$ under the co-adjoint action of the Lorentz group, and the reduced  phase space is the cartesian product of $M$ with this co-adjoint orbit \cite{duval-horvathy-annals}.
After quantization of the  co-adjoint orbit, the quantities $\kappa$ and $\Sigma$ will define the highest weights  in the resulting   representation of the Poincare group \cite{balachandran}.   Different choices of the matrix $\Sigma^{ab}$  lead to different values for the intrinsic spin of the particle. Similarly different choices for the matrix    $\kappa= \kappa^a \gamma_a$ allow us to consider both massive and massless particles within  one formalism. 

If we compute   
\be
[\Sigma,\kappa]= \gamma_a \Sigma^{ab}\kappa_b,
\ee
we see that  $[\Sigma,\kappa]=0$ is equivalent to  $\Sigma^{ab}\kappa_b=0$, and by Lorentz covariance this is  in turn  equivalent to $S_{ab}k^b=0$.  But $[\Sigma,\kappa]=0$  means that  $\Sigma$ lies in the the Lie algebra of the little-group of $\kappa$.  As  $S_{ab}k^b=0$ is a property possessed by the  Dirac angular momentum  $S_{ab} =i \bar\psi \sigma_{ab} \psi$ (see equation (\ref{EQ:nice-condition}))  we  will accept this  little-group property as a  natural constraint  on the spin tensor. In the relativity literature it is  known  as the Tulczyjew-Dixon  condition  \cite{tulczyiew,dixon}.   It is to be contrasted with the rival Mathisson-Pirani \cite{mathisson,pirani} condition $S_{ab}\dot x^b=0$, where 
\be
\dot x^b= e^b_\mu\frac{d x^\mu}{d\tau}.
\ee
Here $\tau$ can be  any  coordinate that parameterizes the space-time trajectory $x^\mu(\tau)$. It does not have to be the proper time.

When  $\lambda$  depends on  $\tau$ we have 
\bea
\frac{d}{d\tau } S_{ab}&=& - {\rm tr}\{[\Sigma, \lambda^{-1}\dot \lambda]\lambda^{-1} \sigma_{ab}\lambda\}\nonumber\\
&=& - {\rm tr}\{[ \lambda \Sigma \lambda^{-1} , \dot \lambda\lambda^{-1} ] \,\sigma_{ab}\}.
\eea
The covariant derivative of $S_{ab}$ along the trajectory  $x^\mu(\tau)$ is therefore given by  
\bea
\frac{D}{D\tau } S_{ab}&\stackrel{\rm def}{=}&\frac{d}{d\tau } S_{ab}-  \left(S_{cb}  \,{\omega^c}_{a\mu} +S_{ac}\,{\omega^c}_{b\mu} \right)\dot x^\mu \nonumber\\
&=&- {\rm tr}\{[\Sigma, (\lambda^{-1}\dot \lambda+\lambda^{-1}(\omega_\mu \dot x^\mu) \lambda)]\lambda^{-1}\sigma_{ab}\lambda\}.\nonumber\\
&=&-{\rm tr}\{[\lambda \Sigma \lambda^{-1},\dot \lambda \lambda^{-1}+\omega_\mu \dot x^\mu]\,\sigma_{ab}\}.
\eea
Similarly, from $k_a=m\,{\rm tr}\{\kappa\,\lambda^{-1}\gamma_a \lambda\}$, we get
\be
\frac{d k_a}{d\tau}=- {\rm tr}\{[ \lambda\kappa\lambda^{-1},\dot \lambda \lambda^{-1}] \gamma_a\}
\ee
and hence
\bea
\frac{D}{D\tau} k_a &=&\frac{d}{d\tau } k_a-  k_c\,{\omega^c}_{a\mu}\dot x^\mu\nonumber\\
&=& -{\rm tr}\{[ \lambda \kappa \lambda^{-1},\dot \lambda \lambda^{-1}+\omega_\mu \dot x^\mu]\gamma_a\}.
\eea

Now we introduce some one-forms that we will use to build the classical action functional for our particle. 
Let  $e^*= {\bf e}^{*a}\gamma_a$ so we  can write $\tilde{\bf e}^{*a}= {[\Lambda^{-1}]^a}_b{\bf e}^{*b}$ as $\tilde e^*= \lambda^{-1}e^*\lambda$. We  use this to write  
\be
k_\mu dx^\mu ={\rm tr}\{\kappa\, \lambda^{-1} { e}^*\lambda\}\stackrel{\rm def}{=}\Omega_1.
\ee
which is  to be considered as a  one-form on the total space $P$, rather than on the base space  $M$.

Next  define 
\be
\tilde \omega= \textstyle{\frac 12} \sigma_{ab} \tilde \omega^{ab}\stackrel{\rm def}{=}  \lambda^{-1} \left(d +\textstyle{\frac 12} \sigma_{ab} \omega^{ab}\right)\lambda=\lambda^{-1}(d+\omega)\lambda.
\ee  
This is again  1-form on the total space of the bundle $\pi:P\to M$. The  $\tilde \omega^{ab}$  are zero  on the horizontal subspace of $H\subset T(P)$  each point on the fibre,  while the $\tilde {\bf e}^{*a}$ are zero  on the vertical subspace of  $V\subset T(P)$.
We use these forms to define 
 \be
 \Omega_2= {\rm tr}\{\Sigma\, \lambda^{-1}(d+\omega)\lambda\}.
 \ee

We take as the action functional 
\be
S[x, \lambda]= \int  \Omega,
\label{EQ:action}
\ee
where 
\be
\Omega= \Omega_1-\Omega_2,
\ee
and the integral is taken along the  curve parameterized by $\tau$.
As shown in \cite{dwivedi-stone}, the  equations of motion are  
\be
i_X d\Omega=0.
\ee
where $X$ is a vector field tangential to  the trajectory in $P$.

To compute $d\Omega_1$ we will assume that the spin connection is torsion free, so that
\be
d{\bf e}^{*a}+  {\omega^a}_b\wedge \ {\bf e}^{*b}=0.
\ee
We can then use 
\be
[\sigma_{ab},\gamma_c]=( \gamma_a \eta_{bc}- \gamma_b \eta_{ac})
\ee
 to see that
 \bea
 d\Omega_1&=&d \,{\rm tr}\{\kappa \,\lambda^{-1} { e}^*\lambda\} \nonumber\\
 &=& -{\rm tr}\{[\lambda\kappa \lambda^{-1}, d\lambda \lambda^{-1}+ \textstyle{\frac 12} \sigma_{ab} \omega^{ab}] \,e^*\}.
 \eea

For  $d\Omega_2$ we need the matrix-valued Riemann curvature tensor  
 \be
d  {\omega} +{\omega}\wedge  {\omega}= \frac 12 ({\frac 12} \sigma_{ab} R^{ab})_{\mu\nu}dx^\mu dx^\nu\equiv R,
\ee
and observe that if $\tilde \omega= \lambda^{-1}(d+ \textstyle{\frac 12} \sigma_{ab} \omega^{ab})\lambda$ we have
\be
d  \tilde {\omega} +\tilde {\omega}\wedge  \tilde {\omega}=\lambda^{-1}( \textstyle{\frac 12} \sigma_{ab}  R^{ab})\lambda\equiv \lambda^{-1}R\lambda.
\ee
Consequently 
\bea
d\Omega_2&=& 
d \,{\rm tr\!}\left\{\Sigma \,\lambda^{-1}\!\left(d+ \textstyle{\frac 12} \sigma_{ab} \omega^{ab}\right)\!\lambda\right\} \nonumber\\
&=&{\rm tr\!}\left\{\lambda \Sigma \lambda^{-1}\,R)\right\} - {\rm tr\!}\left\{\lambda \Sigma\lambda^{-1} \left(d\lambda \lambda^{-1} + \textstyle{\frac 12} \sigma_{ab} \omega^{ab}\right)^2\right\}.\
\eea
We will write $d\lambda \lambda^{-1}+\omega=\tilde \omega_{\rm R}= \textstyle{\frac 12} \sigma_{ab}\tilde \omega^{ab}_{\rm R}$. (The ``R''   because  $\tilde \omega^{ab}_{\rm R}$   includes the  right-invariant Maurer-Cartan form $d\lambda \lambda^{-1}$.) We note that $e^{*a}$ and $\tilde \omega^{ab}_{\rm R}$ are linearly independent and  between them  span  ${\rm T}^*(P)$.  

We can evaluate the contractions $i_Xd\Omega\equiv d\Omega(X)$ by using 
 \bea
 e^*(X)&=& \dot x^a\gamma_a = \dot x^\mu e^{*a}_\mu \gamma_a,\nonumber\\
d\lambda \lambda^{-1}(X) &=& \dot \lambda\lambda^{-1}, \nonumber\\ 
 R(X)&=& -{\textstyle\frac 12} \sigma_{ab} {R^{ab}}_{\mu\nu} dx^\mu \dot x^\nu=  -{\textstyle\frac 12} \sigma_{ab} {R^{ab}}_{\mu\nu} \dot x^\nu e^\mu_a e^{*a},\nonumber\\
 \omega(X)&=& {\textstyle\frac 12} \sigma_{ab} {\omega^{ab}}_\mu \dot x^\mu.
 \eea 
 Here $\dot x^\mu$ denotes  $dx^\mu /d\tau$.
We find that
\bea  
i_X d\Omega_1 &=&- {\rm tr}\{[\lambda \kappa\lambda^{-1},\dot \lambda \lambda^{-1}+ \omega_\mu \dot x^\mu]\gamma_a\}e^{*a}+
{\rm tr}\{ \lambda\kappa \lambda^{-1}[\sigma_{ab}, \gamma_c]\} \dot x^c \tilde \omega^{ab}/2\nonumber\\
 &=&-{\rm tr}\{[\lambda \kappa\lambda^{-1},\dot \lambda \lambda^{-1}+ \omega_\mu \dot x^\mu]\gamma_a\}e^{*a}+
{\rm tr}\{ \lambda\kappa \lambda^{-1}(\gamma_a \eta_{bc}- \gamma_b \eta_{ac}) \dot x^c) {\textstyle \frac 12}  \tilde \omega^{ab}_R\nonumber\\
&=& \left(\frac{D k_a}{D\tau}\right) e^{*a} +(k_a\dot x_b-k_b\dot x_a) {\textstyle \frac 12 }\tilde \omega_R^{ab}
\eea
and 
\bea
i_X d\Omega_2&=&-{\rm tr}\{\lambda \Sigma \lambda^{-1} R_{\mu\nu} \dot x^\nu e^\mu_a\} e^{*a}-
{\rm \tr}\{\lambda\Sigma \lambda^{-1} ,\dot \lambda \lambda^{-1}+\omega_\mu \dot x^\mu]\sigma_{ab}\}  \tilde \omega^{ab}_R/2\nonumber\\
&=&\left(- \frac 12 S_{mn}{R^{mn}}_{\mu\nu} e^\mu_a \dot x^\nu\right)e^{*a} + \left(\frac{D S_{ab}}{D\tau}\right) {\textstyle \frac 12 }\tilde \omega_R^{ab}.
\eea
The contraction  $i_Xd\Omega$ is therefore a position-dependent combination of $e^{*a}$  and $\tilde \omega_R^{ab}$. For it to be  zero, we need   the coefficients of these forms  be separately zero.
Requiring the vanishing of the coefficient of  
$e^{*a}$ yields 
\be
\frac{D}{D\tau}k_c+\textstyle{\frac 12} S_{ab}{R^{ab}}_{\mu\nu} \dot x^\nu e^\mu_c=0.
\label{EQ:MP-momentum}
\ee
Similarly, the vanishing of the coefficient of $\tilde \omega_R^{ab}$ gives 
\be
\frac{D}{D\tau}S_{ab}+ \dot x_a k_b- k_a\dot x_b=0.
\label{EQ:MP-ang-momentum}
\ee
These are the  Mathisson-Papapetrou-Dixon  \cite{mathisson,papapetrou,dixon} equations.  The momentum equation  (\ref{EQ:MP-momentum}) exhibits  a gravitational analogue of the Lorentz force, while  (\ref{EQ:MP-ang-momentum}) expresses  the conservation of  total (spin and orbital) angular momentum. It is well known that to obtain a closed system these two equations have to be supplemented by a condition on the spin  such as our  Tulczyjew-Dixon  condition  $k^aS_{ab}=0$.  It is explained in appendix \ref{SEC:appendix-centroid} that this condition means that $x^\mu(\tau)$ is the worldline of the particle's centre of mass.

Before we proceed  there is  a necessary consistency check. Our entire action principle is  built on the assumption that $k^2=m^2$ is fixed ---  but the RHS  of (\ref {EQ:MP-momentum}) does not immediately seem to ensure that   $k^ak_a$ is  a constant of the motion.  To verify that it is,  we can write $k_a=m u_a$ where $u_au^a=1$. We then contract the both sides of the momentum equation with $v^c=\dot x^c$ and  use the antisymmetry of the curvature  tensor to see that 
\be
m \dot x^a \dot u_a+ \dot m \dot x^au_a = 0.
\ee
Now from
\be
u_a S^{ab}=0
\ee
we get 
\be
u_a \dot u_b\dot S^{ab}=- \dot u_a\dot u_b S^{ab}=0,
\ee
and hence from the angular momentum  equation we find that
\be
0=u_a \dot u_b(k^a\dot x^b-\dot x ^ak^b)=m (\dot u_b \dot x^b - \dot u_b u^b u_a \dot x^a)= m \dot u^b \dot x_b.
\ee
Thus $0=\dot m\, (\dot x^au_a)$ and the mass is indeed a constant of the motion. This constancy   continues when we include a Lorentz force. It would  {\it not\/} survive   were we to  include an explicit magnetic moment. In that case the action would need to be extended  to accommodate a modified  mass-shell condition 
\cite{duval-mass}.

\subsection{The anomalous velocity due to spin}

It is the Mathisson-Papapetrou-Dixon angular-momentum equation (\ref{EQ:MP-ang-momentum}), with its implication that $\dot x^a$ is no longer parallel to $k^a$,  that gives us the anomalous velocity.
From equation (\ref{EQ:MP-ang-momentum})
and the Tulczyjew-Dixon  little-group condition $k^aS_{ab}=0$ we deduce that
\be
-\frac{D k^a}{D\tau}S_{ab}=k^2 \dot x_b -k_b (\dot x \cdot k).
\ee
or
\be
 \dot x_a= \frac{1}{m^2} \left( (k_a  (\dot x \cdot k)+S_{ac} \frac{D k^c}{D\tau}\right).
 \label{EQ:velocity1}
\ee
There are several things that we can do with this result.

Firstly, substituting (\ref{EQ:velocity1})  into   the angular momentum conservation law (\ref{EQ:MP-ang-momentum}) we find 
\be
\frac{D S_{ab}}{D\tau}+\frac 1{m^2}\left(S_{ac} k_b \frac{D k^c}{D\tau}+ S_{cb} k_a \frac{D k^c}{D\tau}\right)=0.
\ee
This  is Fermi-Walker transport of the spin angular-momentum tensor  along  the trajectory  whose tangent vector is $k^\mu/m$ rather than $\dot x^\mu$. Dixon \cite{dixon}  calls this {\it M-transport\/}.
  
Secondly we can find  the  ``anomalous'' correction to the relation between velocity and  momentum. Up to now the parameter $\tau$ was  arbitrary.  The action is reparametrization invariant so $\tau$ does not have to be the proper time. If we  change the parameterization   $\tau\to t$  in such a manner that   the  vielbein component  $\dot x_0$ becomes unity, then the remaining  $\dot x_i$, $i=1,\ldots, d-1$, are the components of the  velocity ``3''-vector in the local Lorentz frame ${\bf e}_a$ The first component of (\ref{EQ:velocity1}) now becomes 
\be
1= \frac{1}{m^2} \left\{(\dot x \cdot k) E+S_{0c} \frac{D k^c}{Dt }\right\},
\ee
or, rearranging, 
\be
(\dot x  \cdot k) = \frac{m^2+\dot k^aS_{a0}}{E},
\ee
where 
\be
 \dot k^a  \stackrel{\rm def}{=}  \frac{D k^a}{Dt}= \frac{d k^a}{dt}+ {\omega^{ab}}_c k_b \dot x^c.
\ee
Again use $i$ and $j$ for space indices, observe that $k^0=E=\sqrt{m^2+ \sum_{i=1}^3k^i k^i}$ gives 
  \be
  \dot k^0 = \frac{\partial k^0}{\partial k^j} \dot k^j = \frac{k^j}{E} \dot k^j  = - \frac {k_j}{E}\dot k^j,
  \ee
  and make  use of the skew symmetry in $a, b$ of the spin connection ${\omega^{ab}}_\mu$.
  We find that
  \be
  \dot x_i = \frac{k_i}{E}+\frac{1}{m^2}\left(S_{ij}- S_{i0}\frac{k_j}{E}- \frac{k_i}{E}S_{0j}\right)\frac{D k^j}{D t}
  \label{EQ:kunzle-anomalous}
 \ee
 
 Equation (\ref{EQ:kunzle-anomalous})  has a familiar structure! It looks just like  the anomalous velocity equation (\ref{EQ:chang-niu}) with 
 \be
 \Omega_{ij}\to \frac{1}{m^2}\left(S_{ij}- S_{i0}\frac{k_j}{E}- \frac{k_i}{E}S_{0j}\right).
 \ee
 Furthermore, the  associated two-form 
\be 
\frac 12 \Omega_{ij} \,dk^i\wedge dk^j=  \frac{1}{2m^2}\left(S_{ij}- S_{i0}\frac{k_j}{E}- \frac{k_i}{E}S_{0j}\right) dk^i\wedge dk^j 
\ee
 looks very much like our matrix-valued covariant Berry-connection  curvature tensor  
 \be
    {\mathfrak F}_{\alpha\beta}=\frac{1}{2m^2}\left(S_{ij}- S_{i0}\frac{k_j}{E}- \frac{k_i}{E}S_{0j}\right)_{\alpha\beta}dk^i\wedge dk^j,
    \ee
which  in   three dimensions is 
 \be
 {\mathfrak F}_{\alpha\beta}=  \frac 1{2m^2}\frac 1 \gamma  \left\{ \frac 1{2} \left( {\bm \sigma}+ \frac{ ({\bf k}\cdot {\bm \sigma}){\bf k}}{m^2(1+\gamma)}\right)_{\alpha\beta}\right\}\cdot (d{\bf k}\times d{\bf k}).
  \ee
The  quantity in braces  is the lab-frame spin of a particle with polarization ${\bf s}={\bm \sigma}/2$.
It is therefore natural to identify the   classical spin angular momentum $S_{ab}$  with  expectation value 
\be
\bar \psi \frac {i}{4}[\gamma_a,\gamma_b] \psi= C^{*\alpha} (S_{ab})_{\alpha\beta} C^\beta
\ee
of the matrix-valued connection  evaluated in  the WKB state $\psi=u_\alpha C^\alpha$. Were we to quantize by integrating  over $\lambda$ in a path integral, we would  expect  $S_{ab}$  to correspond to the operator  $(S_{ab})_{\alpha\beta}$ that acts in the spin-polarization Hilbert space. 
 
\subsection{Return to the Berry connection}

Our  classical action  (\ref{EQ:action}) leads to   dynamical evolution of the elements $\lambda$ of  the non-compact Lorentz group  ${\rm SO}(1,d-1)$.  In  the previous work  \cite{stephanov,stone-dwivedi,dwivedi-stone} the  phase-space was parametrized by ${\bf x}$, ${\bf k}$,  and elements of a {\it compact\/}  rotation group.     We can  connect the apparently distinct formalisms  by  a simple reparametrization of our degrees of freedom.  We factorize each element $\lambda$ as
\be
\lambda=\lambda_k \sigma,
\ee
where $\lambda_k$ is a chosen  $k$-dependent Lorentz transformation that takes us from the reference ${\bf e}_0$ to  momentum $k$,  and $\sigma$ lies  in the little group of ${\bf e}_0$.   For massive particles this  little group is ${\rm SO}(d-1)$.   The two one-forms  composing  the action (\ref{EQ:action})  now become
\be
\Omega_1= k_\mu dx^\mu
\ee
and 
\be
\Omega_2 ={\rm tr}\{\Sigma\, \lambda^{-1}\! \left(d- \textstyle{\frac 12} \sigma_{ab} \omega^{ab} \right)\lambda\}= {\rm tr}\{\Sigma\,  \sigma ^{-1} \!\left(d+ (\lambda^{-1}_k d\lambda_k) -\textstyle{{\frac 12}} (\lambda^{-1}_k\sigma_{ab} \lambda_k)\omega^{ab} \right)\sigma\}.
\ee
and the action $S[x,\lambda]$ becomes $S[x,k ,\sigma]$.
As    $\Sigma$ lies   in the Lie algebra of little group, the trace operation projects the Lorentz Lie-algebra element $\lambda^{-1}_k d\lambda_k$ into the Lie algebra of the little-group. The projected element    $P \lambda^{-1}_k d\lambda_kP\equiv -i{\mathfrak a}_i dk^i$ is   essentially the non-abelian Berry connection  that produces parallel transport on the little group  in the formalism of \cite{stephanov,stone-dwivedi,dwivedi-stone}.   A gauge transformation on this  Berry connection is a change of choice $\lambda_k \to \lambda_k \sigma_k$ for some $k$-dependent element $\sigma_k$ of the little group.  It is ``essentially''  the same connection rather than ``precisely'' the same  because we have $\lambda^{-1}_k d\lambda_k$ rather than $\lambda^\dagger_k  d\lambda_k$. The present parallel transport is therefore  the non-unitary covariant connection that gives rise to Thomas precession.   In \cite{stephanov,stone-dwivedi,dwivedi-stone} we are  considering massless particles,   and the  Berry connection provides unitary parallel transport on the group  ${\rm SO}(d-2)$. Connecting this massless case to our present formalism requires a  more detailed  consideration that we supply in the next section.

\section{Massless particles}
\label{SEC:massless}

When our particles are  massless the situation becomes rather  more complicated. Even in the free case --- no gravity, no electromagnetic field, and hence $\dot k^a=0$ --- the Mathisson-Papapetrou-Dixon angular momentum  equation
\be
\frac{d S_{ab} }{d\tau}+ \dot x_a k_b-k_a\dot x_b=0
\label{EQ:bad-mathisson}
\ee
supplemented by the Tulczyjew-Dixon condition $S_{ab}k^b=0$  fails to have unique a solution.  Suppose  that  $k^2=0$ and $S_{ab}$ satisfies  $S_{ab}k^b=0$, then 
 \be
\tilde S_{ab}=S_{ab} +(k_a S_{pb} -k_b S_{pa})\Theta^p
\label{EQ:spin-ambiguity}
\ee
still  satisfies $\tilde S_{ab}k^b=0$.  Further,  if $S_{ab}$ and $x_a$ satisfy  (\ref{EQ:bad-mathisson}) and  
  we   set  
\be
\quad \tilde x_a= x_a+ S_{pa}\Theta^p,
\label{EQ:wigner-shift}
\ee
then
$\tilde S_{ab}$, ${\tilde x}_a$ are also  a  solution of (\ref{EQ:bad-mathisson})  for any   time-dependent $\Theta^p(\tau)$.  This multiplicity of solutions  is related to
the absence of a well-defined centre of mass, and to  the corresponding difficulty of defining a covariant spin angular-momentum tensor for massless particles.  

That there  is going to be problem in the massless case is signalled by  the factors of $1/m^2$ in our Berry  curvature tensors. Indeed  we expect a   problem defining the  spin angular-momentum tensor  itself: when  $\psi$ is a Dirac spinor of definite chirality, the tensor  $S_{ab}=i \bar\psi [\gamma_a,\gamma_b]\psi /4$  is identically zero.
To understand the spin of massless particles, we need to appreciate Wigner's observation \cite{wigner}  that the little group for massless  particles is the Euclidean group ${\rm SE}(d-2)$,  and not the na{\"i}vely expected  ${\rm SO}(d-2)$.

For massless particles in $d$-dimensional Minkowski space  we can take the reference-momentum   einbein   to be the null-vector 
\be
N^a=\underbrace{(1,0,\ldots, 0,1)}_{d}.
\ee 
The Lie algebra of the  little group of $N^a$ consists of the   $\sigma_{ab}$ with $0<a,b,<d$ that generate  ${\rm SO}(d-2)$, together with 
\be
\pi_a\stackrel{\rm def}{=} N^b\sigma_{ba} = \sigma_{0a}+\sigma_{da} ,\quad 0< a<d.
\ee 
Indeed, we can check that
\be
[\pi_a , N^b\gamma_b]=0 , \quad 0< a<d.
\ee
From 
\be
[\sigma_{ij},\sigma_{mn}]=\eta_{jm}\sigma_{in}  -\eta_{im}\sigma_{jn}-\eta_{jn}\sigma_{im}+\eta_{in}\sigma_{jm}
\ee
we find that
\be
[\pi_a,\pi_b]=0, \quad [\sigma_{ab}, \pi_c]= \eta_{bc} \pi_a- \eta_{ac} \pi_b.
\ee
The $\pi_a$ therefore  behave like translations, and  together with the rotations  generate  the Euclidean group ${\rm SE}(d-2)$.  Wigner argues in  \cite{wigner} that the quantum states of all known particles must  be unaffected by these ``translations.''

For example, consider  the 3+1 massless Dirac equation.   For $N^a=(1,0,0,1)$ we have
\be
\pi_1 =-  \frac 12 \left[\matrix{i\sigma_2&  \sigma_1\cr \sigma_1&i\sigma_2}\right],\quad
\pi_2 = \frac 12 \left[\matrix{i\sigma_1 &  -\sigma_2\cr -\sigma_2&i\sigma_1}\right],
\ee
and both these ``translation'' operators act as zero on  the relevant  positive energy, positive and negaive  chirality, states   
\be
u_+(N)= \left[\matrix{1\cr0\cr1\cr 0}\right], \quad u_-(N)= \left[\matrix{0\cr1\cr0\cr -1}\right].
\ee

We can obtain a general  null-momentum  $k^a= (|{\bf k}|, {\bf k})= e^s (1, {\bf n})$  by applying to $N^a$ a rapidity-$s$ boost parallel to the ${\bf e}_3$  direction, and then  a rotation that takes ${\bf e}_3$ to the unit vector ${\bf n}$. In the Dirac representation, this procedure is implemented by   
\be
\lambda_k = \exp\{-i\phi \Sigma_3\}  \exp\{-i\theta \Sigma_2\}  \exp\{sK_3\},
\ee
where $\theta$ and $\phi$  are the polar angles of the direction of the three-momentum ${\bf k}$, and 
\be
\Sigma_i = \frac 12 \left[\matrix{\sigma_i&0 \cr 0 &\sigma_i}\right], \quad K_i  =\frac 12 \left[\matrix{\sigma_i&0 \cr 0 &-\sigma_i}\right],
\ee
are respectively the rotation and boost generators.
The resulting covariantly-normalized spinor positive chirality spinor  is $u_+({\bf k})=\lambda_k u_+(N)$ is 
\be
u_+({\bf k}) = e^{s/2} \left(\matrix{\chi\cr \chi}\right),
\ee
where
\be
\chi({\bf k})= \left(\matrix{\cos (\theta/2)\cr e^{i\phi}\sin(\theta/2)}\right).
\ee
 The Dirac-equation eigenstates are therefore safely indifferent  to  any  Wigner translations in $\lambda_k\to \lambda_k \sigma$. 

The same is not true of the  classical angular momentum tensor $S_{ab}= \tr\{\Sigma \lambda^{-1} \sigma_{ab}\lambda \}$.  If we 
replace
\be
\lambda \to \lambda \exp\left(\sum_{i=1}^{d-2} \theta^i \pi_i \right)\, 
\ee
then we have a transformation 
\be
\delta_{{ \Theta}}: S_{ab} \to S_{ab}+ (k_a S_{pb}-k_bS_{pa}) \Theta^p
\label{EQ:wigner-spin-action}
\ee
where $\Theta^p = {\Lambda^p}_i \theta^i$ and $k^a= {\Lambda^a}_b N^b $.   Thus $S_{ab}$ is affected by the unphysical Wigner translations in the same manner as  in (\ref{EQ:spin-ambiguity}).  The Wigner-translation operation differs   from that in  (\ref{EQ:spin-ambiguity}), however,  in that the parameter $\Theta^p$ in (\ref{EQ:spin-ambiguity}) is arbitrary but the parameter in (\ref{EQ:wigner-spin-action})  must satisfy $\Theta^p k_p=0$.  This  constraint follows from the relation $\Theta^p = \Lambda^p_i \theta^i$, and is necessary for two successive translations with parameters $\Theta_1^p$ and $\Theta_2^p$ to be equivalent to one with parameters $\Theta_1^p+ \Theta_2^p$. In particular, a  transformation that is allowed by (\ref{EQ:spin-ambiguity}) but not by (\ref{EQ:wigner-spin-action})  is given by $\Theta_0^p\equiv  (-E^{-1},0,\ldots 0)$. It takes 
$$
\delta_{{\Theta}_0}: S_{ab}\to \left(S_{ab}- \frac{k_a}{E} S_{0b} - S_{a0}\frac {k_b}{E}\right).
$$ 
In other words it takes the spin tensor and projects it to  Pauli-Lubansky tensor.    Any subsequent Wigner translation leaves the  Pauli-Lubansky tensor invariant. 
This tensor  therefore captures  the  physically significant part of the spin angular momentum.   

A   Wigner translation, when combined with the translation $x_a\to x_a + S_{pa}\Theta^p$, leaves the free action invariant even for time-dependent $\Theta^p(\tau)$. The Wigner translation group must  therefore be regarded as a {\it gauge invariance\/}  \cite{balachandran-atre}.   The gauge group is slightly larger than just the Wigner translations because the action on $x_a$ is not abelian. Again requiring  $\Theta^p k_p=0$, we find that  
\be
[\delta_{{ \Theta}_2}, \delta_{{\Theta}_1}] x_a = 2  \,\Theta^p_1 \Theta^q_2 S_{pq}\, k_a
\ee
This means that  translations  $x_a \to x_a+ \varepsilon k_a$ must also be included in the gauge group of the free action \cite{balachandran-atre}. 

Being gauge variant, the position of   $x_a$ of the particle is not an observable. This seems like a disaster for any mathematical model  that claims to describe the motion of a particle.  All is not lost, however. What has happened is that a massless particle has no rest frame and therefore no observer-independent centre of mass. As explained in appendix \ref{SEC:appendix-centroid}, it still has well-defined mass {\it  centroids\/}, but the location of these centroids depends on the reference  frame of the observer.

In our massless action, we  are still free to fix a gauge,  and so pin down a position for the particle.  A  natural gauge choice is to factorise $\lambda = \lambda_k
 \sigma$ where $\sigma$ is chosen to be an element of ${\rm SO}(d-2)$. In other words, we deliberately excluding the problematic Wigner translations  from  our action. 
Once we do this the free action becomes 
\be
\int\left( k_\mu dx^\mu - {\rm tr}\{ \Sigma \,\sigma^{-1}(d+\lambda^{-1}_k d\lambda_k)\sigma\}\right), 
\ee
and this is of the same form as the action in \cite{stephanov,stone-dwivedi,dwivedi-stone} where the internal spin degree of freedom lives only in the rotation part of the little group. 
For example, in 3+1 dimensions we write
\be
\sigma= \exp\{i\Sigma_3 \varphi\}
\ee
and 
\be
\lambda=\lambda_k \sigma = \exp\{-i\phi \Sigma_3\}  \exp\{-i\theta \Sigma_2\}  \exp\{sK_3\} \exp\{i\varphi \Sigma_3\}.
\ee
If we take  take $\Sigma= J\Sigma_3/4$ then 

\bea
\Omega_2 &=& \textstyle{ \frac 14} J {\rm tr}\{\Sigma_3 \lambda^{-1} d\lambda\}\nonumber\\
&=&  \textstyle{ \frac 14} J{\rm tr}\{\Sigma_3 \sigma^{-1}(d+  \lambda_k^{-1} d\lambda_k)\sigma\}\nonumber\\
&=& iJ{\rm tr} \left( d\varphi -  \cos \theta\, d\phi\right)\nonumber
\eea
The $d\varphi$ is total derivative and does not affect the equation of motion. The $iJ \cos \theta\, d\phi $ term is precisely the Berry phase for a spin $J$ particle. Our  action therefore reduces to that in \cite{stephanov}.    

In general dimensions the gauge fixed action gives the anomalous velocity  of the lab-frame centroid in terms of Wigner-translation invariant Pauli-Lubanski tensor. 
 \be
  \dot x_i = \frac{k_i}{E}+\frac{1}{E^2}\left(S_{ij}- S_{i0}\frac{k_j}{E}- \frac{k_i}{E}S_{0j}\right)\dot k^j.
  \ee
 In the massless case the  Pauli-Lubanski tensor  not only has vanishing time components (as does the massive case) but  is also perpendicular to the space components of the momentum. This condition is the higher-dimensional analogue of the spin being slaved to the momentum.  

 The gauge-fixing  is frame-dependent, and consequently the  action is no longer manifestly Lorentz covariant. For complete covariance we need to allow $\lambda$ to be any Lorentz transformation matrix --- not only one that omits the Wigner translations.  When we make a Lorentz transformation, we must therefore make a corresponding gauge transformation so as to restore the non-covariant gauge choice in the new frame. The gauge transformation involves the spacetime-translation in (\ref{EQ:wigner-shift}), and this translation corresponds to the   relocation of the   lab-frame mass centroid defined  in (\ref{EQ:centroid-shift}).  This shift accounts for (in the free theory at least) the  unusual Lorentz transformation uncovered  in \cite{son3}.  The  shift is not just a mathematical artifact: the   position of the energy  centroid of of a circularly-polarized light beam is frame-dependent \cite{aiello,bliokh-nori}, and the energy-centroid  is where a photon detector at rest  in this reference  frame would locate the beam. 

There is a fly in the ointment, however.
The Wigner gauge invariance is violated by  interactions. Once $\dot k_a$ is non zero we find that the angular momentum conservation equation changes into  
\bea
\frac{d \tilde S_{ab} }{d\tau}+ \dot {\tilde x}_a k_b-k_a\dot {\tilde x}_b&=&(\dot k_a S_{pb} -\dot k_b S_{pa})\Theta^p\nonumber\\
&=&((x_a-\tilde x_a)\dot k_b- \dot k_a (x_b-\tilde x_b)).
\label{EQ:torque}
\eea
What has happened is that, with  a non-zero net force, the  external  torque depends on the point about which moments are taken.
The  non-zero  right hand side of (\ref{EQ:torque}) is the torque about the new particle location $\tilde x_a$ due to the force acting at the old particle position $x_a$. 

Once we are no longer allowed to make  gauge transformations,  the gauge-fixed theory and  manifestly covariant theory are no longer exactly equivalent. As a consequence exact Lorentz invariance has been lost in the gauge-fixed theory. This may seem  unsatisfactory, but it is to be expected.  There are two related reasons. Firstly the   proof cited in  appendix   \ref{SEC:appendix-centroid}, that the angular momentum  of an extended body defined by (\ref{EQ:ang-mom-tensor-def}) is actually a Lorentz  tensor depends crucially on there being no external force on the body. When we make a Lorentz transformation,  the time-slice integral   samples different epochs  in the body's history, and  the history-dependent  momentum acquisition  spoils the tensor property. This fact necessarily causes a problem for any point-particle approximation to the extended body unless the body is very compact and the external  force small.   The force being small is also a necessary condition for the validity of  the adiabatic approximation which is a prerequisite for  Berry transport. The  adiabatic approximation also depends   on there being a large difference in energy between the $\pm |E|$ states. The point-particle actions used in \cite{stephanov,stone-dwivedi,dwivedi-stone} are therfore  only applicable to particles with a large $E=|{\bf k}|$ --- but a Lorentz transformation can take a large $E=|{\bf k}|$ particle to one with arbitrarily small $E$.  Therefore only we expect Lorentz  invariance only under suitably ``small''  transformations.


\section{Discussion}
\label{SEC:discussion}

We have seen that for massive particles  Berry-phase-containing equations of motion such as  (\ref{EQ:lorentz}) and (\ref{EQ:chang-niu}) can be the 3-dimensional reduction of a   manifestly covariant equation  of motion for the  particle's center of mass. The same is not true for massless particles. In the  absence of a rest frame in which to define the centre of mass,  the best we can do is derive an equation of motion for the lab-frame energy centroid of  the particle, and when the massless particle is spinning the position of this  energy centroid is observer-dependent.  

We can  understand physically why the spin angular-momentum plays a central role in the anomalous velocity.   A spinning object  of mass $m$ and acted on by a force ${\bf F}$ possesses  a   {\it hidden momentum\/}  \cite{wald,costa}  of
  \be
  {\bf P}_{\rm hidden}^{\rm spin}\approx  -\frac{{\bf S }\times {\bf F}}{mc^2}.
  \ee
  (We have restored  the factors of $c$ to emphasize that this is a relativistic effect.) Therefore the  total momentum of the body is given by 
  \be
  {\bf P}_{\rm tot} \approx  m\dot {\bf x} -\frac{{\bf S }\times {\bf F}}{mc^2}.
   \ee
   Identifying $  {\bf P}_{\rm tot} $ with ${\bf k}$, and ${\bf F}$ with $\dot {\bf k}$,  gives us
   \be
   \dot {\bf x}\approx  \frac{{\bf k}}{m} +\frac{{\bf S }\times \dot {\bf k}}{m^2c^2}.
   \ee
 Now,  at low speed, and taking into account that $\dot x_a=-\dot x^a$ and  $k_a=-k^a$,  our anomalous velocity equation  (\ref{EQ:kunzle-anomalous})  reduces to 
  \be
  \dot {\bf x}= \frac{{\bf k}}{m} +\frac {{\bf S}\times {\bf \dot k}}{m^2c^2}, 
  \ee
so anomalous velicity is precisely accounted for by the hidden-momentum.
       
   When  we include the torque produced by a magnetic moment ${\mathcal M_{ab}}$, the angular-momentum conservation equation becomes \cite{souriau}   
     \be
   \frac{D}{D\tau}S_{ab}+\dot x_a k_b-k_a\dot x_b= {F_a}^c {\mathcal M}_{cb} -{\mathcal M}_{ac}{F^c}_b.
   \ee
   (To get an explicit magnetic moment  into the  the action principle, we should  modify the mass-shell condition so that $p^2 =m^2+eg S_{\mu\nu}F^{\mu\nu}/2$,  
   \cite{duval-mass}).
    The   $g$ factor is defined by  
   \be
   {\mathcal M}_{ab} = \frac{ge}{2m}S_{ab},
   \ee
   and  if we approximate (by ignoring terms higher in $S$) 
   \be
   \dot k_b= eF_{bc}\dot x ^c\sim eF_{bc} k^c/m,
   \ee
   then 
   we find that our anomalous velocity equation is replaced by 
   \be
  \dot x_i = \frac{k_i}{E}+\left(1-\frac g2\right)\frac{1}{m^2}\left(S_{ij}- S_{i0}\frac{k_j}{E}- \frac{k_i}{E}S_{0j}\right)\dot k^j.
  \ee
 Again we can understand this {\it via\/} a  hidden momentum.   An accelerating   magnetic dipole has a hidden momentum  of
  \be
  {\bf P}_{\rm hidden}^{\rm EM}= \frac{{\bm  \mu}\times {\bf E}}{c^2}
  \ee
  \cite{coleman,vaidman}, so
  \be
  {\bf P}_{\rm tot} = m\dot {\bf x} + \frac{{\bm \mu}\times {\bf E}}{c^2}-\frac{{\bf S }\times {\bf F}}{mc^2}.
  \ee
  Once  we set  $\mu = ge/m$ and $e{\bf E}=\dot {\bf k}$, we find that 
  \be
    {\bf k} = m\dot {\bf x} + \left(\frac g 2-1\right) \frac{1}{mc^2} {\bf S}\times \bf \dot {\bf k},  
  \ee
  or
  \be
  \dot {\bf x}=\frac{{\bf k}}{m} +\left(1-\frac g2\right) \frac{1}{m^2c^2} {\bf S}\times \bf \dot {\bf k},
  \label{EQ:moment-velocity}
  \ee
  which is again consistent with the Mathisson-Papapetrou-Dixon equation modified to include the effect of the magnetic moment.  That there is no anomalous velocity correction when $g=2$ is also a conclusion  in \cite{horvathy_g-2}.
  
The result in (\ref{EQ:moment-velocity}) does {\it not}   coincide with the wave packet calculation in \cite{niu-review}.    In  \cite{niu-review}  the anomalous velocity is entirely  accounted for by the electromagnetic hidden momentum. There is no sign of the  spin hidden momentum that is  intimately connected with our  Thomas precession curvature. This  discrepancy is presumably  due to the position ``${\bf x}$'' in the massive Mathisson-Papapetrou-Dixon equations being the  centre of mass extracted from moments of the energy momentum tensor. The position ``${\bf x}$''  in \cite{niu-review} is the centre of charge or probability density   of the wave packet. Since a Lorentz-boosted magnetic moment acquires an electric-dipole moment,  the charge centre will move away from the mass-centroid in a velocity-dependent manner  and this momentum-dependendent shift will also contribute to $\dot {\bf x}$. Whether this shift completely  explains the discrepancy  requires further study.

 \section{Acknowledgements}   This work was supported by the National Science Foundation under grant number NSF DMR 13-06011. In the course of this work MS has exchanged many useful emails with Peter Horv\'athy and Christian Duval.  In particular we thank them for sharing with us early drafts of 
 \cite{duval-horvathy-chiral}  that covers many of the topics in the present paper  --- but from a different  yet  complementary  perspective. MS would also like to the Konstantin Bliokh for sanity preserving  discussions about the observer dependence of the location of light rays, and for drawing our attention to \cite{bliokh-nori}. We also thank the authors of \cite{son3} for sending us an early version  of their work. 

\appendix

 \section{Spinors, polarization and  spin}
 \label{SEC:appendix-spin}

In three dimensions we may take the Dirac gamma matrices to be 
\be
\gamma^0= \left[\matrix {1&0\cr 0& -1}\right], \quad \gamma^a=\left[\matrix {0&{ \sigma_a}\cr -{ \sigma_a}&0}\right], \quad \alpha^i =-\gamma_0 \gamma_i=\gamma^0\gamma^i=\left[\matrix {0&{ \sigma_i}\cr { \sigma_i}&0}\right].  
\label{EQ:u-alpha-covariant}
\ee
The  eigenspinors  with Lorentz-covariant normalization  $\bar u_\alpha  u_\beta =-\bar v_\alpha v_\beta =\delta_{\alpha\beta}$,  $\bar u_\alpha  v_\beta =\bar v_\alpha  u_\beta =0$
can be taken to be 
\bea
u_\alpha({\bf k}) &=& \frac{1}{\sqrt{2m(E+m)}} \left[\matrix{ (E+m)\chi_\alpha \cr  ({\bm \sigma}\cdot {\bf k} )\chi_\alpha}\right],\nonumber\\
v_\alpha({\bf k}) &=&  \frac{1}{\sqrt{2m(|E|+m)}} \left[\matrix{ ({\bm \sigma}\cdot {\bf k} ) \chi_\alpha \cr  (|E|+m)\chi_\alpha}\right],
\label{EQ:chidef}
\eea
where  $\chi_\alpha$  are  the unit  two-spinors $\chi_1=(1,0)^T$ and $\chi_2=(0,1)^T$. The label $\alpha$ on $u_{\alpha}$ is  therefore that of  the spin  in the rest frame of the particle, where
\be 
u_{\alpha}\to \left[\matrix{ \chi_\alpha \cr  0}\right].
\ee
The spin  in the particles's rest frame is usually called the ``polarization,'' and is a more transparent  quantity to work with than the lab-frame spin \cite{rosen}.  

Define the spin generators 
\be
\Sigma_{\mu\nu}= \frac i 4[\gamma_\mu,\gamma_\nu]
\ee
and assemble the spatial parts into  into a spin three-vector ${\bm \Sigma}= (\Sigma_{23},\Sigma_{31},\Sigma_{12})$ where
\be
{\bm \Sigma}=\frac 12 \left[\matrix{{\bm \sigma}&0 \cr 0 &{\bm \sigma}}\right].
\ee
We can now evalaute 
\bea
 u^\dagger_\alpha {\bm \Sigma}u_\beta 
&=& \frac{1}{2}\,\chi_\alpha^\dagger\! \left({\bm \sigma} +\frac{1}{m^2}\frac{{\bf k}({\bf k}\cdot {\bm \sigma})}{(1+\gamma)}\right)\!\chi_\beta\nonumber\\
&=& \frac{1}{2}\left({\bm \sigma} +\frac{1}{m^2}\frac{{\bf k}({\bf k}\cdot {\bm \sigma})}{(1+\gamma)}\right)_{\alpha\beta}\nonumber\\
&=& \frac{1}{2}\left({\bm \sigma} +\frac{\gamma^2 {\bm \beta}({\bm \beta}\cdot {\bm \sigma})}{(1+\gamma)}\right)_{\alpha\beta},
\label{EQ:lab-spin}
\eea
where ${\bm \beta}= {\bf v}={\bf k}/E$  is the three-velocity, and $\gamma=(1-|{\bm \beta}|^2)^{-1/2}$.  

The physical meaning of the combination of ${\bm \sigma}$'s  in parentheses in (\ref{EQ:lab-spin}) 
can be understood by defining   a  spin four-vector $(S^0, {\bf S})$ that  takes the value  $(0,{\bf s})$ particle's rest frame. Then, by performing a Lorentz transformation, we   find that the corresponding  lab-frame  components  are given by 
\bea
{\bf S}&=& {\bf s}  +\frac{\gamma^2 {\bm \beta}({\bm \beta}\cdot {\bf s})} {(1+\gamma)}\nonumber\\
S^0 &=& \gamma{\bm \beta}\cdot {\bf s}= {\bm \beta}\cdot {\bf S} .
\eea We see that   $u^\dagger _\alpha {\bm \Sigma}u_{\beta}$ coincides with the Lorentz transform of the matrix elements  of the operator that measures the polarization ${\bf s}$.

Alternatively, we can define the Pauli-Lubansky spin four-vector operator   
\be
{\mathfrak S}^\kappa=\frac 12  \epsilon^{\kappa\mu\nu\lambda}\Sigma_{\mu\nu}\left( \frac {k_\lambda}{m}\right), \quad \epsilon^{1230}=1,
\ee
that   reduces to $(0,{\bm \Sigma})$  in the particle's rest frame where $k^\mu=(m,{\bf 0})$.  
Its  three  space components are
\be
{\mathfrak S}^i= \frac 12 \gamma \,\epsilon^{ijk} \left(\Sigma_{jk} -\frac{k_i}{E} \Sigma_{0k} - \Sigma_{j0}\frac{k_k}{E}\right),
\ee
and the   time component is 
\be
{\mathfrak S}^0 = \gamma({\bm \beta}\cdot {\bm \Sigma}) = {\bm \beta}\cdot {\mathfrak  S}.
\ee
Because the  the matrix elements   ${\bar u}_\alpha \Sigma_{\mu\nu}u_\beta$  transform as a tensor,  and the matrix elements of the  space components $\Sigma_i$ and ${\mathfrak S}_i$ coincide  in the particle's rest frame, we  must have that
\be
\frac 1 {\gamma} u^\dagger_\alpha { \Sigma}_iu_\beta = {\bar u}_\alpha\left\{ \frac 12  \epsilon^{ijk} \left(\Sigma_{jk} -\frac{k_i}{E} \Sigma_{0k} - \Sigma_{j0}\frac{k_k}{E}\right)\right\}u_\beta,
\label{EQ:spin-identity}
\ee
as can be confirmed by direct calculation. The left hand side of (\ref{EQ:spin-identity}) comprises  the matrix elements of the 3-spin operator in a plane-wave beam normalized to  one particle per unit volume in the lab frame.  A physical interpretation of the right hand side is provided in appendix \ref{SEC:appendix-centroid}. 

Although the Pauli-Lubansky four-vector can only be defined in 3+1 dimensions,  the identity 
\be
\frac 1{\gamma} u^\dagger_\alpha \Sigma_{ij} u_\beta = \bar u_\alpha \left(\Sigma_{ij}- \frac{k_i}{E}\Sigma_{0j} - \Sigma_{i0}\frac{k_j}{E}\right)u_\beta,
\label{EQ:spin-current}
\ee
is true in all dimensions.
Eq (\ref{EQ:spin-current})   follows from setting $\lambda=0$, $\mu=i$, $\nu=j$ in the the covariant identity
\be
\frac{1}{2} \bar u_\alpha \{\gamma_\lambda, \Sigma_{\mu\nu}\}u_\beta =\frac 1 m  \bar u_\alpha\left( k_\lambda \Sigma_{\mu\nu}- k_\mu \Sigma_{\lambda \nu}- \Sigma_{\mu\lambda} k_\nu\right)u_\beta.
\label{EQ:t-identity}
\ee
In turn, equation   (\ref{EQ:t-identity})  holds  because    the right- and left-hand sides are both  totally antisymmetric tensors whose components coincide in the rest frame of the particle.  With the $\lambda$ index raised, the  left-hand side of (\ref{EQ:t-identity})  comprises the matrix elements of the spin-current tensor ${S^{\lambda}}_{\mu\nu}$.

 \section{General properties of Dirac spinors and matrix elements}
 \label{SEC:appendix-general}
 
 We collect some  properties  of the solutions $u_\alpha(k)$, $v_\alpha(k)$  to the plane-wave Dirac equation in any space-time dimension. 
The positive energy eigenvectors $u_\alpha$, $\bar u_\alpha\equiv u^\dagger_\alpha \gamma_0$,  satisfy
\bea
(\gamma_\mu k^{\mu}-m)u_\alpha&=&0,\nonumber\\
\bar v_\alpha (\gamma_\mu k^{\mu}-m)&=&0.
\label{EQ:dirac-u}
\eea
The corresponding negative energy eigenvectors  $v_\alpha(k)$,  $\bar v_\alpha\equiv v^\dagger_\alpha \gamma_0$,  obey 
\bea
(\gamma_\mu k^{\mu}+m)v_\alpha&=&0,\nonumber\\
\bar v_\alpha (\gamma_\mu k^{\mu}+m)&=&0.
\label{EQ:dirac-v}
\eea
In both cases the momenta lie  on the positive energy mass shell. $k^2=m^2$,  $k^0>0$.
We impose  the Lorentz  covariant  normalization $\bar u_\alpha  u_\beta =-\bar v_\alpha  v_\beta =\delta_{\alpha\beta}$, $\bar u_\alpha v_\beta=\bar v_\alpha u_\beta=0$. The completeness relation is therefore 
\bea
{\mathbb I}&=& u_\alpha \bar u_\alpha- v_\alpha\bar v_\alpha\nonumber\\
&=&\, \,\,\Lambda_+\,\,+\,\,\Lambda_-, 
\eea
where the projection operators  are  
\bea
\Lambda_+&=& 
 \frac{1}{2m}(m+\gamma_\mu k^\mu)=\phantom- u_\alpha\bar u_\alpha\nonumber\\
 \Lambda_-&=&\frac{1}{2m}(m-\gamma_\mu k^\mu)=- v_\alpha\bar v_\alpha.
\eea 

By varying the equation $\bar u_\alpha(\gamma_\mu k^{\mu}-m)u_\beta=0$ and making use of the normalization conditions,   we find the  4-current matrix elements 
\be
\bar u_\alpha \gamma^\mu u_\beta = \bar v_\alpha \gamma^\mu v_\beta = \delta_{\alpha\beta}\,k^\mu/m\equiv  \delta_{\alpha\beta} V^\mu.
\label{EQ:V-from-u}
\ee
Here  $V^\mu = \gamma(1, {\bm \beta})$ is the 4-velocity derived from the group velocity ${\bm \beta}= \partial k^0/\partial {\bf k}$.

Similarly, by  varying (\ref{EQ:dirac-u}), (\ref{EQ:dirac-v}), we find that 
\bea
\bar v_\alpha\,\delta u_\beta &=& \frac 1{2m} \delta k^\mu (\bar v_\alpha \gamma_\mu u_\beta),\nonumber\\
\delta \bar u_\alpha\, v_\beta &=& \frac 1{2m} \delta k^\mu (\bar u_\alpha \gamma_\mu v_\beta).
\label{EQ:dirac-vary}
\eea
We must keep $k^\mu$  on the mass-shell;  consequently  the $\delta k^\mu$ are not independent.

Now consider the covariant Berry connection   ${\mathfrak a}_{\alpha\beta}=i\bar u_\alpha du_\beta$. From  $\bar u_\alpha du_\beta+ d\bar u_\alpha u_\beta=0$, (\ref{EQ:dirac-vary}),  and the completeness relation, we find that the corresponding curvature is given by 
\bea
{\mathfrak F}_{\alpha\beta} &\stackrel{\rm def}{=}& (d{\mathfrak a}-i {\mathfrak a}^2)_{\alpha\beta}
\nonumber\\
&=& (id\bar u_\alpha du_\beta -  d\bar u_\alpha u_\gamma \bar u_\gamma du_\beta)\nonumber\\
&=& i(d\bar u_\alpha u_\gamma \bar u_\gamma du_\beta- d\bar u_\alpha v_\gamma \bar v_\gamma du_\beta -d\bar u_\alpha u_\gamma \bar u_\gamma du_\beta)\nonumber\\
&=& - i(d\bar u_\alpha v_\gamma)( \bar v_\gamma du_\beta)\nonumber\\
&=& - i(\bar u_\alpha \gamma_\mu v_\gamma)( \bar v_\gamma \gamma_\nu u_\beta) dk^\mu\wedge dk^\nu/4m^2\nonumber\\
&=&i(\bar u_\alpha \gamma_\mu  \gamma_\nu u_\beta) dk^\mu\wedge dk^\nu/4m^2\nonumber\\
&=& i(\bar u_\alpha [\gamma_\mu , \gamma_\nu ]u_\beta) dk^\mu\wedge dk^\nu/8m^2\nonumber\\
&=& \frac 1 {2m^2} (S_{\mu\nu})_{\alpha\beta} \, dk^\mu\wedge dk^\nu. 
\eea

Another set of covariant  matrix elements   are the  
\be
(S_{\mu\nu})_{\alpha\beta}= \bar u_\alpha \Sigma_{\mu,\nu}u_\beta.
\ee
that occur in (\ref{EQ:spin-identity}), (\ref{EQ:spin-current}) and (\ref{EQ:t-identity}).
They play the role of the components  of a covariant  angular momentum tensor.
We find directly  from the Dirac equation, that they obey
\be
k^\mu (S_{\mu\nu})_{\alpha\beta}=0. 
\label{EQ:nice-condition}
\ee

\section{Covariant Berry transport is Fermi-Walker transport}
\label{SEC:appendix-fermi-walker}

Here we  show that if  we expand  $\psi= u_\beta C^\beta$ and $\bar \psi= C^{*\alpha}\bar u_\alpha$ then  the  covariant parallel transport  of the coefficients $C^\alpha$  leads to the angular momentum tensor  
\be
S^{\mu\nu}\equiv \frac{i}{4} \bar\psi \Sigma^{\mu\nu} \psi
\ee
being  Fermi-Walker transported along the trajectory.

Berry transport of the  $C^\alpha$ means  that 
\be
\delta C^\alpha =- (\bar u_\alpha \delta u_\beta) C^\beta.
\ee
The states $u_\alpha(k)$ themselves change with $k^\mu$ so that
\be
\delta u_\beta = u_\alpha\bar u_\alpha \delta u_\beta- v_\alpha\bar v_\alpha \delta u_\beta.
\ee
Putting these two results together we have
\be
\delta(u_\beta C^\beta)= -v_\alpha(\bar v_\alpha\delta u_\beta) C^\beta,
\ee
and
\be
\delta (C^{*\alpha} \bar u_\alpha)= -C^{*\alpha} (\delta \bar u_\alpha v_\beta) \bar v_\beta.
\ee

We now  use  the formul{\ae}\   (\ref{EQ:dirac-vary})   for  $ (\delta \bar u_\alpha v_\beta) $ and $(\bar v_\alpha\delta u_\beta) $ to find
\bea
4i \delta S^{\mu\nu}&=&  C^{*\alpha}\delta \bar u_\alpha v_\rho\bar v_\rho [\gamma^\mu,\gamma^\nu] \psi-\bar \psi [\gamma^\mu,\gamma^\nu] v_\sigma\bar v_\sigma \delta u_\beta C^\beta,\nonumber\\
&=& \frac{\delta k^\lambda}{2m}\left( \bar\psi \gamma_\lambda  v_\rho\bar v_\rho [\gamma^\mu,\gamma^\nu] \psi- \bar\psi  [\gamma^\mu,\gamma^\nu] v_\sigma\bar v_\sigma \gamma_\lambda \psi\right),\nonumber\\
&=& +\frac{\delta k^\lambda}{2m} (\bar\psi \gamma_\lambda  v_\rho\bar v_\rho \gamma^\mu u_\sigma \bar u_\sigma \gamma^\nu \psi- (\mu\leftrightarrow \nu))\nonumber\\
&&-\frac{\delta k^\lambda}{2m} (\bar\psi \gamma_\lambda  v_\rho\bar v_\rho \gamma^\mu v_\sigma \bar v_\sigma \gamma^\nu \psi- (\mu\leftrightarrow \nu))\nonumber\\
&& +\frac{\delta k^\lambda}{2m} (\bar\psi  \gamma^\mu u_\rho\bar u_\rho \gamma^\nu v_\sigma\bar v_\sigma \gamma_\lambda \psi- (\mu\leftrightarrow \nu))\nonumber\\
&&-\frac{\delta k^\lambda}{2m} (\bar\psi  \gamma^\mu v_\rho\bar v_\rho \gamma^\nu v_\sigma\bar v_\sigma \gamma_\lambda \psi- (\mu\leftrightarrow \nu)).
\eea
We can simplify by using the current matrix elements to get
\bea
 \delta S^{\mu\nu}&=& \frac{i}{4 m^2} \left( k^\nu \delta k^\lambda \bar
\psi [\gamma^\mu,\gamma_\lambda] \psi+  k^\mu \delta k^\lambda \bar\psi [\gamma_\lambda, \gamma^\nu]\psi\right)\nonumber\\
&=& \frac{1}{m^2}   \left(  k^\nu \delta k^\lambda {S^\mu}_\lambda + k^\mu \delta k^\lambda {S_\lambda}^\mu\right)
\eea
Thus we have found that 
\be
\frac{ \partial S^{\mu\nu}}{\partial \tau} +  \frac{1}{m^2}   \left(  k^\nu \frac{\partial  k^\lambda}{\partial \tau} {S^\mu}_\lambda + k^\mu \frac{\partial k^\lambda}{\partial \tau} {S_\lambda}^\nu\right)=0.
\label{EQ:FW!}
\ee
At the level of  WKB, where we see no anomalous velocity,  we have  $V^\mu= k^\mu/m$. Consequently   (\ref{EQ:FW!}) is 
\be
\frac{ \partial S^{\mu\nu}}{\partial \tau} +    V^\nu \frac{\partial  V^\lambda}{\partial \tau} {S^\mu}_\lambda + V^\mu \frac{\partial V^\lambda}{\partial \tau} {S_\lambda}^\nu=0.
\label{EQ:FW2}
\ee
Given that $V_\mu S^{\mu\nu}=(k_\mu/m) S^{\mu\nu}= 0$,  equation (\ref{EQ:FW2}) is the statement that  $S^{\mu\nu}$ is being Fermi-Walker transported.

 \section{Centroids and the centre of mass} 
 \label{SEC:appendix-centroid} 

We  review  some standard material on centroids and centres of mass of extended bodies that should apply to  wave-packets of Dirac particles.
We work  in flat space and suppose there are no external forces.   Our  extended body therefore possesses  a  conserved and compactly supported symmetric energy-momentum tensor
\be
\partial_\mu T^{\mu\nu}=0, \quad T^{\mu\nu}=T^{\nu\mu}.
\ee
Let $x_{\rm A}^\mu$ be a space-time event, $\Sigma$ a spacelike surface,  and define the angular momentum of the body about $x_{\rm A}$ by
\be
M^{\mu\nu}_{\rm A} = \int_\Sigma\left\{(x^\mu-x_{\rm A}^\mu)T^{\nu\gamma}- (x^\nu-x_{\rm A}^\nu)T^{\mu\gamma}\right\}d\Sigma_\gamma
\label{EQ:ang-mom-tensor-def}
\ee
then (\cite{MTW} page 161) $M^{\mu\nu}_{\rm A} $ is a tensor, and independent of the choice of   $\Sigma$.

We now choose a lab  frame and, with $i,j$ running over space indices only, we  define the energy and three-momentum of the body to be
\be
E= \int _{x^0=t}T^{00}d^3x, \quad p^i =   \int_{x^0=t} T^{i0}d^3x,
\ee
respectively.
We also define the {\it mass-centroid\/} $X^i_{\rm L}$ in the lab frame by  
\be
\left\{\int_{x^0=t} T^{00}d^3x \right\}\,X^i_{\rm L}= \int_{x^0=t} x^iT^{00}d^3x.
\ee

Now 
\be
\partial_t  \int_{x^0=t} T^{00}d^3x =  \int_{x^0=t} \partial_0 T^{00}d^3x = - \int _{x^0=t}\partial_j T^{j0}d^3x=0,
\ee
and 
\be
\partial_t  \int_{x^0=t} x^i T^{00}d^3x =  \int _{x^0=t} x^i\,\partial_0 T^{00}d^3x =   -\int_{x^0=t}  x^i\,\partial_j T^{j0}d^3x=  \int _{x^0=t}\delta^i_j \,T^{j0}d^3x= p^i.
\ee
So, differentiating its definition with respect to $t$, we read off that the ordinary three-velocity of the centroid is 
\be
\dot {\bf X}_{\rm L}= {\bf p}/E.
\ee

Now take  $x^\mu_{\rm A}$  to be point in the  $x^0=t$ surface. Then
\bea
M^{i0}_{\rm A} &=& \int_{x^0=t} \left\{(x^i-x_{\rm A}^i)T^{00}- (x^0-x_{\rm A}^0)T^{i0}\right\}d^3x\nonumber\\
&=& (X^i_{\rm L}-x^i_{\rm A})E.\nonumber
\eea
(The second term on the right in the first line  is zero because $x^0- x^0_{\rm A}$  is zero everywhere in the integral.)
Thus $M^{i0}_A$ is zero when ${\rm A}$ is the centroid in the lab frame.  If we replace the lab frame with an inertial  frame having four-velocity $V^\mu$ we have that 
$M^{\mu\nu}_{\rm A} V_\nu=0$ if and only if ${\rm A}$ is the mass centroid in that frame.

Define the {\it centre of mass\/} $X^i_{\rm CM}$ to be the mass-centroid in the frame where $p^i=0$, and the {\it intrinsic angular momentum\/} $S^{\mu\nu}$ to be the angular momentum about the centre of mass.  Thus $S^{\mu\nu}p_\nu=0$ and we automatically have the Tulczyjew-Dixon  condition.

Now, looking back at the definition of angular momentum, we see that if we change reference points we have
\be
M^{\mu\nu}_{\rm A}+  x_{\rm A}^\mu p^\nu-  x_{\rm A}^\nu p^\mu= M^{\mu\nu}_{\rm B}+  x_{\rm B}^\mu p^\nu-  x_{\rm B}^\nu p^\mu.
\ee 
Let us take $\x_{\rm A}={\bf X}_{\rm CM}$ and $\x_{\rm B}= {\bf X}_{\rm L}$ to be the centroid in the lab frame. Then
\be
S^{\mu\nu}_{\rm A}+  X_{\rm CM}^\mu p^\nu-  X_{\rm CM}^\nu p^\mu= M^{\mu\nu}_{\rm L}+  X_{\rm L}^\mu p^\nu-  X_{\rm L}^\nu p^\mu, 
\ee
and
\be
 S^{\mu\nu}+ (X_{\rm CM}^\mu - X_{\rm L}^\mu )p^\nu-  (x_{\rm CM}^\nu-X_{\rm L}^\nu) p^\mu=M^{\mu\nu}_{\rm L}.
 \ee
 The lab-frame centroid condition gives us $M^{i0}_{\rm L}=0$, and  we have  $(X_{\rm CM}^0 - X_{\rm L}^0 )=0$,   so
 \be
 S^{0\nu}-  (X_{\rm CM}^\nu-X_{\rm L}^\nu)E=0. 
 \ee
 We write this as 
 \be
 (X_{\rm CM}^\nu-X_{\rm L}^\nu)= \frac{S^{0\nu}}{E} , \qquad \left(= \frac {1}{E^2} S^{\nu i}p_i \right)
\label{EQ:centroid-shift}
 \ee
 and  find
 \be
 M^{\mu\nu}_{\rm L}= \left(S^{\mu\nu}- S^{\mu0}\frac{p^\nu}{E}- \frac{p^\mu}{E}S^{0\nu}\right).
 \ee
 Thus we have   a physical  interpretation of the Pauli-Lubansky  spin-tensor  components  that appears many times in this paper. It is the intrinsic angular momentum about the lab-frame centroid. 
 

 \end{document}